\documentclass[useAMS,usenatbib,usedcolumn]{mn2e}

\usepackage{graphicx} 
\usepackage{times}

\title[Time variation in the low frequency spectrum of Vela-like pulsar B1800$-$21]{Time variation in the low frequency spectrum of Vela-like pulsar B1800$-$21}
\author[Basu et al.]{Rahul Basu$^1$\thanks{e-mail: rahul@astro.ia.uz.zgora.pl}, Karolina Ro\.zko$^1$, Wojciech Lewandowski$^1$, Jaros\l{}aw Kijak$^1$, Marta Dembska$^2$\\
1. Janusz Gil Institute of Astronomy, University of Zielona G\'ora, ul. Szafrana 2, 65-516 Zielona G\'ora, Poland \\
2. DLR Institute of Space Systems, Robert-Hooke-Str. 7, 28359 Bremen, Germany\\
}
\begin{document}



\maketitle

\label{firstpage}

\begin{abstract}
We report the flux measurement of the Vela like pulsar B1800$-$21 at the low 
radio frequency regime over multiple epochs spanning several years. The 
spectrum shows a turnover around the GHz frequency range and represents a 
typical example of gigahertz-peaked spectrum (GPS) pulsar. Our observations
revealed that the pulsar spectrum show a significant evolution during the 
observing period with the low frequency part of the spectrum becoming 
steeper, with a higher turnover frequency, for a period of several years before 
reverting back to the initial shape during the latest measurements. The 
spectral change over times spanning several years requires dense structures, 
with free electron densities around 1000--20000~cm$^{-3}$ and physical 
dimensions $\sim$220 AU, in the interstellar medium (ISM) traversing across the
pulsar line of sight. We look into the possible sites of such structures in the
ISM and likely mechanisms particularly the thermal free-free absorption as 
possible explanations for the change.
\end{abstract}


\section{Introduction}
The pulsar radio emission is nonthermal in nature and characterized by a steep 
power law spectra. Generally, the observed spectra can be modelled with a 
negative spectral index of about $-$1.8~in the frequency regime between 0.1 - 
10 GHz \citep{maron2000}. The gigahertz-peaked spectrum (GPS) observed in a 
handful of cases, where a turnover is seen in the spectra around 1 GHz 
\citep{kijak2007}, provide a deviation from the general pulsar population. 
These sources are mostly present in exotic environments with associated dense 
HII regions or Pulsar Wind Nebulae (PWNe). The GPS nature is believed to 
originate due to interaction of the pulsar emission with the surrounding medium
but is still poorly understood. In order to discover more cases of GPS pulsars 
and gain better understanding of this peculiar phenomenon systematic studies 
have been carried out over the past few years \citep{kijak2011a,kijak2011b,
kijak2013,dembska2014,dembska2015a,dembska2015b}. 

The majority of GPS pulsars are characterized by high dispersion measure (DM 
$>$ 200~pc cm$^{-3}$), which may very well be a coincidence of the specialized 
environments that harbour them. The conventional flux measurement techniques 
which rely on estimating the baseline level of the pulse profile may be 
affected by interstellar scattering which results in smearing of the pulse 
across the entire period. In some cases, when the pulse broadening time is a 
significant fraction of the pulse period (30\% or more) one can see a 
relatively sharp pulse, but at the same time the extended scattering tail may 
obscure the real baseline level, which leads to an underestimation of the
pulsar flux. For pulsars with DMs in 200-300~pc cm$^{-3}$ range this usually 
happens between 300 and 600 MHz \citep{lewandowski2013,lewandowski2015a}. This 
leads to a somewhat pseudo correlation between high DM and GPS pulsars 
\citep{kijak2007,kijak2011b} where serious underestimation of the flux at lower
frequencies for high DM pulsars may give rise to an inverted spectra. The 
interferometric imaging technique provide a more robust measurement of the 
pulsar flux owing to the baseline lying at zero level thereby reducing errors 
made during the baseline subtraction. \citet{dembska2015a}, D15 hereafter, used
the imaging technique to measure the flux of six high DM pulsars at 610 MHz to 
explore their suspected GPS nature. Their studies revealed that in four of the 
six cases the measured flux was underestimated due to scattering effects and 
the corrected flux followed a normal power law spectra. In pulsar B1823$-$13 
they confirmed the presence of GPS characteristics. However, the case of PSR 
B1800$-$21 defied expectations as the interferometric flux measurements turned 
out to be significantly lower than the earlier measurements using conventional 
methods. This gave rise to various questions regarding the spectral nature of 
this pulsar. There was also a possibility of the two measurement techniques 
being incompatible and errors during the measurement processes.

In this work we carry out a detailed study of the spectrum in PSR B1800$-$21 
spanning multiple frequencies and different epochs. Interferometric flux 
measurements at 325 MHz and 610 MHz at two widely separated epochs were carried
out. In addition a complete low frequency spectrum for the pulsar was 
constructed by using data at 325, 610 and 1280 MHz, separated over short 
intervals, simultaneously with a phased array and an interferometer. In each of
these cases the radio spectra of PSR B1800$-$21 were constructed and revealed 
GPS characteristics. PSR B1800$-$21 is a young pulsar similar to the well known
Vela pulsar and is associated with a very interesting environment. The pulsar 
lies within the W30 complex which comprises of a supernova remnant (SNR) and a
number of compact HII regions \citep{kw90,fo94}. This makes the pulsar a well
suited candidate for exhibiting GPS characteristics. We carry out a detailed 
comparison of our measured spectra with previously reported values 
\citep[][D15]{lorimer1995,kijak2011b} and look into the possible implications.

\section{Observations and data analysis}

\begin{table}
\resizebox{\hsize}{!}{
\begin{minipage}{80mm}
\caption{Observing details for PSR B1800$-$21.}
\centering
\begin{tabular}{cccc D{,}{\pm}{3.3}}
\hline
 & & & &  \\
Obs Date & Frequency & Obs Mode & Calibrator & \multicolumn{1}{c}{Cal Flux}\\
 & ({\footnotesize MHz}) &  &  & \multicolumn{1}{c}{({\footnotesize Jy})}\\
\hline
28 Dec, 2013 & 610 & Int & 1822$-$096 & 6.2,0.4 \\
07 Jan, 2014 & 610 & Int & 1822$-$096 & 6.2,0.4 \\
 & & & & \\
03 Jan, 2015 & 325 & Int & 1822$-$096 & 2.7,0.2 \\
17 Jan, 2015 & 325 & Int & 1822$-$096 & 3.4,0.3 \\
 & & & & \\
15 Aug, 2015 & 610 & Int/Ph & 1714$-$252 & 4.7,0.3 \\
20 Aug, 2015 & 325 & Int/Ph & 1822$-$096 & 3.3,0.3 \\
23 Aug, 2015 & 1280 & Int/Ph & 1751$-$253 & 1.0,0.1 \\
29 Aug, 2015 & 610 & Int/Ph & 1714$-$252 & 4.5,0.3 \\
03 Sep, 2015 & 1280 & Int/Ph & 1714$-$252 & 2.4,0.2 \\
14 Sep, 2015 & 325 & Int/Ph & 1714$-$252 & 5.0,0.3 \\
 & & & & \\
\hline
\multicolumn{5}{l}{\footnotesize Int---Interferometric ~~~~~~~ Ph---Phased Array }\\
\end{tabular}
\label{tabobs}
\end{minipage}}
\end{table}

We used the Giant Metrewave Radio Telescope (GMRT) for our studies which is an
interferometric array consisting of 30 dishes in a Y-shaped array spread out 
over a region of $\sim$27 km. The GMRT is uniquely positioned to be used 
simultaneously as an interferometer and a phased array due to alternate routes 
of signal post processing for these two modes. We carried out observations 
which enabled us to compare the flux measurements using the two different 
techniques and verify previous flux values. The GMRT operates in the metre 
wavelengths at five distinct frequency bands around 1.4 GHz, 610 MHz, 325 MHz, 
240 MHz and 150 MHz with a bandwidth of 33 MHz at each frequency spread over 
256 frequency channels. For our studies we have used the three frequencies 
1280, 610 and 325 MHz during one or more epochs to measure the flux in pulsar 
B1800$-$21. The lower frequencies were not used for our studies as the pulsar 
was expected to have an inverted spectrum and the flux values would be below 
detection limit at these frequencies. The observations were carried out over 
three different epochs between December, 2013 and September, 2015 as detailed 
in Table \ref{tabobs}. Each of the observations were carried out for more than 
an hour and at each epoch the pulsar was observed twice separated by at least a
week to account for variations due to scintillation. 

\begin{figure}
   \begin{flushleft}
   \includegraphics[scale=0.7]{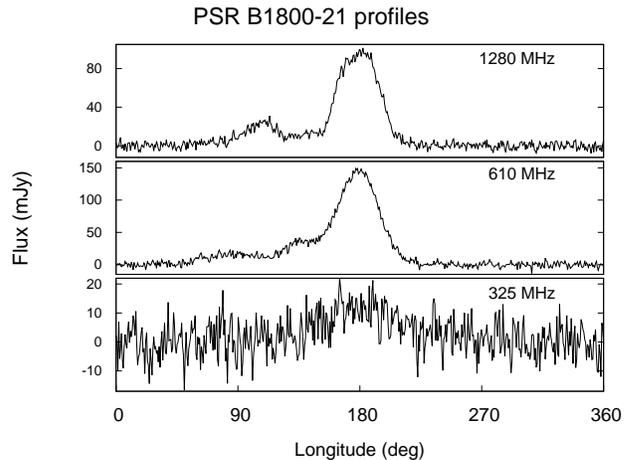}
   \caption{Profile of the pulsar B1800$-$21 measured using GMRT at the three 
radio frequencies 1280 MHz, 610 MHz and 325 MHz. The observations were carried
out between August and September, 2015. The pulsar flux decreases considerably
at the lowest frequency where the profile also exhibits scattering.}
  \label{prof}
   \end{flushleft}
\end{figure}

We used standard observing schemes where the flux calibrator 3C286 was recorded
at the start of each observing run and strategically placed phase calibrators 
(see Table~\ref{tabobs}) interspersed at regular intervals. The Astronomical 
Image Processing System (AIPS) was used for imaging as detailed in D15. The 
flux scale of the calibrator 3C286 was set using the latest Perley-Butler 
(2013) estimates (in AIPS) and was used to calculate the flux of the different 
phase calibrators as shown in Table~\ref{tabobs}. In the phased array mode the 
signal from each of the antennas (around 20 nearby antennas were used) were 
coadded to produce a time series data with resolution 122 microseconds. The 
antennas were phase aligned at regular intervals on the phase calibrator before
being coadded for maximum signal to noise measurements. The data were 
dedispersed and a folded profile for each observation was obtained using the 
topocentric pulsar period (Figure \ref{prof}). We used publicly available 
softwares TEMPO\footnote{http://www.atnf.csiro.au/research/pulsar/tempo} and the
ATNF pulsar database\footnote{http://www.atnf.csiro.au/people/pulsar/psrcat/} 
\citep{man2005} for determining the different parameters of the pulsar 
B1800$-$21 during our observing runs. All the calibrators were observed in the
on-off mode, where data were recorded with the antennas pointing away from the 
source before recording on the source. The resulting shift in the count level
from the background gave a scaling factor for the measured counts to the flux
levels in jansky. The average pulsar flux from the pulse profiles were 
calculated after subtracting the baseline and suitably scaling with the scaling
factor. Finally, the average flux during all the epochs for each observing 
frequency were determined for the two observing modes and shown in Table 
\ref{tabflux}.

\section{Results}
\subsection{Interferometer Vs Phased Array}
One of the primary goals of our studies was to verify the compatibility of the
two different techniques used to measure the pulsar flux. This was prompted by 
the results reported in D15 where the measured flux of B1800$-$21 using 
interferometric method was considerably lower than the previously reported 
measurements at 610 MHz \citep{kijak2011b}. We have carried out simultaneous 
flux measurements using the two techniques at three different frequencies as 
shown in Table \ref{tabflux}. The measurements are identical at all the three 
frequencies, within measurement errors, showing the two methods to be mutually 
consistent in determining the flux of pulsars. It is still possible that one or 
more of the reported flux in the literature is affected by observing errors, 
but our studies confirm that there is no measurement bias due to the observing 
technique employed. We look into the implications of the spectral nature of the
pulsar B1800$-$21 with emphasis on the physical conditions in the following 
sections. It is worth noting that Table \ref{tabflux} shows the interferometric
flux measurements to be more accurate with smaller errors compared to the 
phased array values. This may be due to a number of reasons which we list below
:\\
1. In the phased array the number of antennas used ($\sim$ 20) is less than the
antennas used for the interferometric measurements ($\gid$ 27). This is because
in the phased array mode the signals from each antenna is phase aligned at 
regular intervals on a calibrator before being coadded for higher signal to 
noise detection. The further apart the antennas are the faster the alignment 
is lost and as a result only nearby antennas are used to save time on phasing.
The interferometer on the other hand records all the correlation between 
antenna pairs where the phase alignment is done offline. This requires longer
computation time during analysis but all the available antennas can be used for
the measurements.\\
2. The interferometer records the correlation between two antennas as a result
of which the baseline due to background is at zero level. This results in more 
robust measurements free from the errors made during baseline subtraction while
estimating the scaling factor and pulsar profile in the phased array. A 
significant drawback of the phased array flux measurement arises in the case 
of highly scattered pulsars with long tails rendering baseline levels 
indeterminate. The interferometric measurements provide the only secure way of 
estimating flux in these sources (see D15).\\
3. Finally, the most important advantage using the interferometer is that the 
instrumental and atmospheric gain fluctuations can be corrected on very short 
time scales using self-calibration of the interferometric data. These 
corrections are determined by flux densities of constant and bright background 
sources in the field and considerably reduces the noise levels.\\
The number of data points recorded at any instant is considerably larger in an 
interferometer compared to a phased array. Hence, the time resolution of an 
interferometer is significantly lower which is an major drawback in 
conventional pulsar studies. However, this is not a primary requirement for 
flux estimation where the interferometric technique provide a superior 
alternative to other traditional means.

\begin{table}
\resizebox{\hsize}{!}{
\begin{minipage}{80mm}
\caption{Flux Measurement for PSR B1800$-$21.}
\centering
\begin{tabular}{cc D{,}{\pm}{3.3} D{,}{\pm}{3.3}}
\hline
 & & & \\
 Observation Date & Frequency & \multicolumn{1}{c}{Intfr Flux} & \multicolumn{1}{c}{Ph-Arr Flux}\\
 & ({\footnotesize MHz}) & \multicolumn{1}{c}{({\footnotesize mJy})} & \multicolumn{1}{c}{({\footnotesize mJy})} \\
\hline
Dec, 2013 - Jan, 2014 & 610 & 9.2,0.7 & \multicolumn{1}{c}{---} \\
 & & & \\
Jan, 2015 & 325 & 3.4,0.3 & \multicolumn{1}{c}{---} \\
 & & & \\
Aug, 2015 - Sep, 2015 & 325 & 3.4,0.2 & 2.2,2.6 \\
Aug, 2015 - Sep, 2015 & 610 & 12.8,0.9 & 14.3,4.8 \\
Aug, 2015 - Sep, 2015 & 1280 & 13.9,1.0 & 12.8,3.5 \\
 & & & \\
\hline
\end{tabular}
\label{tabflux}
\end{minipage}}
\end{table}

\subsection{Spectral nature of PSR B1800$-$21}
\begin{figure}
   \begin{flushleft}
   \includegraphics[angle=-90]{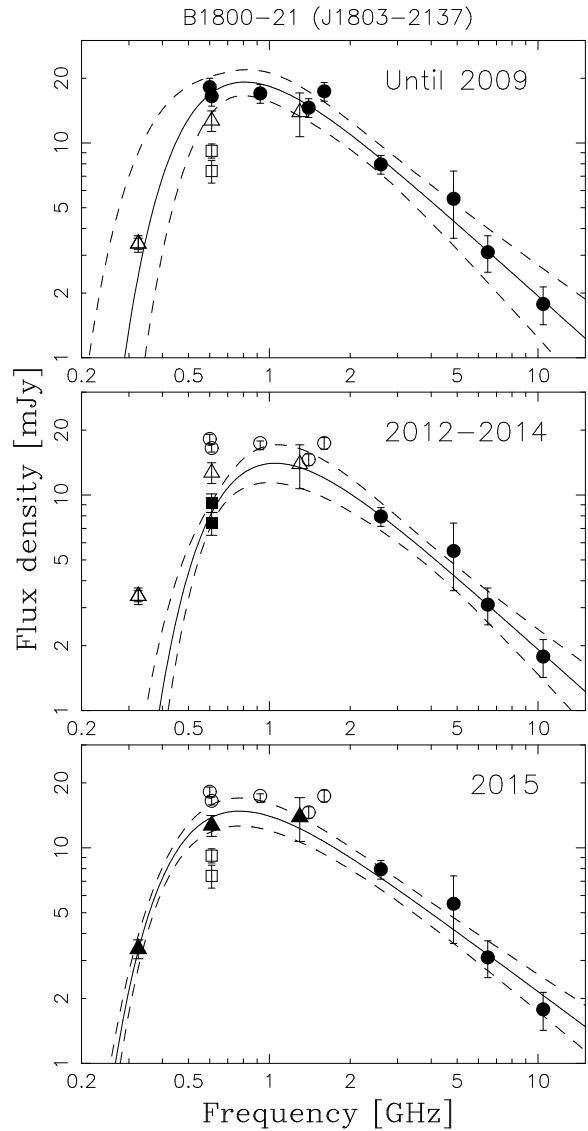}
   \caption{The radio spectra for the pulsar B1800$-$21 measured during 
different observing epochs. The circular points represent the flux values from
the literature measured before 2009, the square points show the measurements
during the period between December, 2012 and January, 2014. The measurements
carried out after January, 2015 are shown as triangular points. The solid line 
represents the fitted model during each epoch of observation and the dashed 
lines correspond to the $1 \sigma$ envelope for all possible fits (the fitting 
procedure is described in the main text). For each case the spectral fits 
were performed only using the measurements denoted by filled symbols with the 
empty symbol measurements shown only for comparison. All three spectral fits 
included the high frequency data (2 GHz and above) from before 2009, to help us
constrain the intrinsic pulsar spectrum (since at these frequencies the effects
of thermal absorption should be negligible). The GPS characteristic is visible 
during each epoch of observation.}
  \label{spectra}
   \end{flushleft}
\end{figure}
The observations reported in this paper were spread out over a period of about 
two years and were motivated by the significantly lower flux measurements of 
the pulsar B1800$-$21 reported in D15, observed on December, 2012, compared to 
the previous epochs. Our measurements between December, 2013 and January, 2014 
at 610 MHz were consistent with the flux values reported in D15 within 
measurement errors, though the actual values in 2013-2014 (9.2$\pm$0.9 mJy, 
Table \ref{tabflux}) were slightly higher than in 2012 values (7.4$\pm$0.9 mJy 
in D15). The low frequency flux from January, 2015 onwards have been much 
higher and closer to the earlier reports. We believe that our results in 
conjunction with D15 show an evolution of the low frequency spectrum of the 
pulsar B1800$-$21~in recent times where the low frequency part of the 
spectrum became steeper around 2012-2014 which has once again reverted to the 
original shape (as in 2009 and earlier) in 2015. In Figure \ref{spectra} the 
pulsar spectra is shown at three different epochs, before 2009 using archival 
measurements, between 2012 and 2014 and finally from January 2015 onwards. In 
all three cases the spectra exhibit GPS characteristics. The thermal free-free 
absorption in the intervening medium first proposed by 
\citet{kijak2011a,kijak2013} to explain GPS behaviour and further extended by 
\citet{lewandowski2015} has been utilized to model the spectrum in each case 
with the spectral fits and errors during fitting shown as solid and broken 
lines respectively, the details of which are presented in the next section. The
GPS behaviour is characterized by the turnover in the pulsar radio spectra 
around the GHz frequency range. The turnover frequency is seen to change from 
0.80 GHz before 2009 to 1.05 GHz between 2012-2014 which changed to 0.77 GHz in
2015. Though there have been studies reporting the change in the pulsar period,
dispersion measure, etc. at short timescales we believe our results are the 
first such reported case of spectral change taking place within timescales of 
several years.

\subsection{The fitting procedure}
In this section we describe the modelling of the pulsar spectrum based on the 
fundamental formulations of the radiative transfer equation. 
%
We hypothesise that the GPS behaviour and its evolution in PSR B1800$-$21 
is explained by the thermal free-free absorption in the intervening  medium.
Assuming a simple power-law nature for the intrinsic pulsar spectrum 
($I_{\nu} = A~\nu^{\alpha}$) and a simplified model of optical
depth \citep{RyLa79,Wil2009} the observed pulsar spectrum is modelled as: 
\begin{equation}\label{eqfit}
S(\nu) = A ~ \left( \frac{\nu}{10 \mathrm{GHz}}\right)^{\alpha} ~ e^{-B~\nu^{-2.1}} ,
\end{equation}
where $A$ is the amplitude (i.e. the pulsar intrinsic flux at 10~GHz), $\alpha$ 
the intrinsic spectral index of the pulsar and $B = 0.08235 \times 
T_{\mathrm{e}}^{-1.35}~\mathrm{EM}$, $T_{\mathrm{e}}$ being the electron 
temperature, $\nu$ the observing frequency and EM the Emission Measure. 
The best fit to the measured spectra were obtained using the 
Levenberg-Marquardt least squares algorithm, with $A$, $\alpha$ and $B$ as free 
parameters of the fit. The uncertainties in the fitted parameters were 
calculated using $\chi^2$ mapping. The results of our spectral modelling for 
all three epochs are shown in Table~\ref{tabfit} (along with the normalized 
$\chi^2$). 
\begin{table}
\resizebox{\hsize}{!}{
\begin{minipage}{90mm}
\caption{The parameters obtained by modelling of the spectrum of PSR~B1800$-$21
for three different epochs (see also Fig.~\ref{spectra}).}
\centering
\begin{tabular}{ccccc}
\hline
 & & & \\
 Obs Epoch & \multicolumn{1}{c}{$A$} & \multicolumn{1}{c}{$B$} & \multicolumn{1}{c}{ $\alpha$} & $\chi^2$ \\
 & \multicolumn{1}{c}{(mJy)} & \multicolumn{1}{c}{({\footnotesize K$^{-1.35}$ pc cm$^{-6}$})} & & \\
\hline
 & & & & \\
 Until 2009 & $1.94^{+0.76}_{-0.70}$ & $0.34^{+0.18}_{-0.17}$ & $-1.12^{+0.24}_{-0.28}$ & 1.33  \\
 & & & &  \\
 2012-2014 & $1.93^{+0.46}_{-0.45}$ & $0.59^{+0.17}_{-0.17}$ & $-1.12^{+0.22}_{-0.25}$ & 0.81 \\
  & & & & \\
 2015 & $2.16^{+0.44}_{-0.41}$ & $0.258^{+0.035}_{-0.035}$ & $-0.93^{+0.12}_{-0.12}$ & 0.73 \\
 & & & & \\
\hline
\end{tabular}
\label{tabfit}
\end{minipage}}
\end{table}

\subsection{Interpretation of spectral changes}
In this section we analyse the constraints on the physical parameters of the 
absorber based on our modelling and the measured value of the dispersion 
measure (DM = $233.99 \pm 0.05$ pc cm$^{-3}$, see \citealt{hlk+04}). 

In principle the fraction of the observed DM contributed by the absorber 
is unknown. However, for our calculations (following the prescription of 
\citealt{raj15}) we assumed 50\% of the DM is contributed by the absorber. If 
we further assume uniform electron distribution ($n_{\mathrm{e}}$) inside the 
absorber with width $s$ along the line of sight, then:
\begin{equation}\label{eq1}
\Delta \mathrm{DM} = n_{\mathrm{e}} \times s,
\end{equation}
\begin{equation}\label{eq2}
\mathrm{EM} = n_{\mathrm{e}}^2 \times s.
\end{equation}
For a given width of the absorber $s$ the electron density is given by Eq.
\ref{eq1} which can be used in turn to estimate the EM using Eq. \ref{eq2}. 
Combining this information with the value of the $B$ parameter obtained from 
the spectral fit (see Eq.~\ref{eqfit}) results in an estimate of the electron 
temperature needed to explain the amount of absorption observed in the 
spectrum. 

It is evident that we are dealing with a degenerate problem with insufficient 
information from the pulsar spectra to constrain the physical parameters of the 
absorber. However, following \citet{lewandowski2015} and \citet{raj15} we point
out three distinct categories of possible absorbers in the interstellar medium 
(ISM) : 
\begin{enumerate}
 \item dense filaments in SNR with typical $s = 0.1$~pc,
 \item cometary shaped tail in PWN with $s = 1.0$~pc,
 \item and HII region with $s = 10.0$~pc.
\end{enumerate}
As a first step towards understanding the turnover in the spectra we used the
initial spectra (before 2009) and for each of the above three configuration
of the absorber determined the physical parameters (the temperature and 
electron density) required for explaining the $B$ parameter reported in Table 
\ref{tabfit} (first line). The physical parameters for each of the three 
categories are shown in Table~\ref{tabcon}, titled `Initial absorber'.

One of the main outcomes of our studies have been the discovery of the change 
in the pulsar spectra (see Figure \ref{spectra}) between 2012 and 2014. The 
absorption during this period was much stronger as evidenced by the higher 
value of the $B$ parameter in Table~\ref{tabfit}. We suggest that this excess 
absorption was caused by an additional absorber that moved across the line of 
sight. In this scenario the overall optical depth is a sum of the contributions
from the initial absorber and the additional component. Using Eq.~\ref{eq1} and
\ref{eq2} along with the difference between the values of $B$ parameter 
estimated for the initial (until 2009) and the increased absorption period 
(between 2012 and 2014) we can constrain the physical parameters of this 
additional absorber. The size of the additional absorber can be estimated using
the pulsar's transverse velocity and the duration over which the spectral 
change was seen. Our observations were inadequate to determine the actual 
timescales for the spectral change, however the minimum and maximum duration 
was five years and one year, respectively. We used a mean duration of three 
years for the change and $v = 347^{+57}_{-48}$ km s$^{-1}$ \citep{bri06} to 
estimate the physical size of the additional absorber to be $s\sim0.001$~pc 
(approximately $220$ AU). We assumed that the physical dimensions of the 
additional absorber were identical along the transverse direction and the line 
of sight. We lack any measurement of the DM during the period between 2012 and 
2014 which once again presented a degenerate problem with both the temperature 
and electron density of the additional absorber being unknown. We have once 
again considered three situations where such conditions can arise in the ISM 
and assumed the typical temperatures as shown in Table \ref{tabcon}, lower part
titled `Additional absorber'. The required electron densities for each case is 
calculated and shown in the table.

In the next section we explore the consequence of our estimates of the physical
parameters of the absorber on the pulsar emission and particularly their 
environmental implications.
\begin{table}
\resizebox{\hsize}{!}{
\begin{minipage}{80mm}
\caption{The constraints on the physical parameters of the absorbing medium for both the initial and additional absorption (see text for detailed explanation).}
\centering
\begin{tabular}{lccc}
\hline
 & \multicolumn{1}{c}{s} & \multicolumn{1}{c}{n$_{\mathrm{e}}$} & \multicolumn{1}{c}{T} \\
 & \multicolumn{1}{c}{({\footnotesize pc})} & \multicolumn{1}{c}{({\footnotesize cm$^{-3}$})}  & \multicolumn{1}{c}{({\footnotesize K})}\\
\hline
 Initial absorber & & & \\
 & 0.1 & 1170 & 2200 \\
 & 1.0 & 117 & 400 \\
 & 10.0 & 12 & 73 \\
\hline
 Additional absorber & & & \\
 & 0.001 & 16657 & 5000 \\
 & 0.001 & 3520 & 500 \\
 & 0.001 & 1188 & 100 \\
 \hline
\end{tabular}
\label{tabcon}
\end{minipage}}
\end{table}

\section{Discussion}
The pulsar B1800$-$21 is young (age $\sim$ 16 kyr) with a high spin down energy
loss (\.E = 2.2$\times$10$^{36}$ erg s$^{-1}$) and is similar to the well known
Vela pulsar which also shows a turnover in the spectrum around 600 MHz 
\citep{sie73}. PSR B1800$-$21 has a very distinct environment and is believed 
to be associated with the W30 complex \citep{kw90,fo94}. The W30 complex is a 
roughly spherical structure consisting of SNR G8.7$-$0.1 and a large number of 
compact HII regions in and around it. The pulsar is located in the southwest 
direction near the edge of the SNR. Detailed proper motion studies by 
\citet{bri06} have ruled out a direct association between PSR B1800$-$21 and 
the supernova remnant, with the pulsar moving approximately towards the center 
of the SNR. High resolution X-ray analysis using the \textit{Chandra X-Ray 
observatory} have revealed the pulsar to be associated with a compact PWN 
\citep{kar07}. We do not have a detailed estimate of the conditions in the
immediate vicinity of the pulsar which makes it difficult to constrain the
spectral fitting. However, we used information about the different structures 
seen around the pulsar to constrain the physical properties of the absorber 
responsible for the GPS phenomenon as shown in Table \ref{tabcon}. The 
proximity of the pulsar with the SNR is the motivation for invoking the 
filamentary structures seen in SNRs as a likely source for the observed GPS 
phenomenon \citep{kijak2011b}, the first condition in Table \ref{tabcon} 
(Initial absorber, $n_{\mathrm{e}}$ = 1170~cm$^{-3}$, T = 2200 K). The physical
parameters of the filament required for the observed turnover in the spectrum 
(around 800 MHz) is similar to the measured values in certain cases 
\citep{koo07}. It is noteworthy that given the small size of the filaments it 
is rare for these structures to be moving across the line of sight of a pulsar.
The second condition in Table \ref{tabcon} (Initial absorber, $n_{\mathrm{e}}$ 
= 117~cm$^{-3}$, T = 400 K) corresponds to the GPS phenomenon associated with 
an asymmetric PWN as has been detected around PSR B1800$-$21 \citep{kar07}. The
GPS phenomenon driven by bow-shock nebula has been investigated by 
\citet{lewandowski2015} where the phenomenon is affected by the relative 
orientation of the direction of pulsar emission and the shape of the bow shock.
The physical parameters in such a situation is consistent with our estimates. 
In principle we do not know how an asymmetric PWN will affect the pulsar 
spectrum, but we can not exclude the possibility of GPS phenomenon arising in 
these systems. Finally, the presence of HII regions near the pulsar prompted us
to investigate the third condition in Table \ref{tabcon} (Initial absorber, 
$n_{\mathrm{e}}$ = 12~cm$^{-3}$, T = 73 K). It should be noted that several of 
the HII regions around the PSR B1800$-$21 show a turnover in the spectrum near 
the GHz frequency range which can be attributed to thermal absorption 
\citep{kw90}. It is possible to have large HII regions with electron densities 
around 10--100~cm$^{-3}$ but the temperatures associated with them are much 
higher (1000--10000 K) \citep{tsa03}, which is contrary to our expectations for
the observed GPS phenomenon. Thus we discount these parameters as the physical 
characteristics of a potential absorber. The above discussions highlight that 
the likely condition giving rise to the GPS phenomenon in this pulsar is the 
absorption in filamentary structures in the surrounding SNR which may also 
account for the changes seen in the pulsar spectrum, as discussed below. 
However, we cannot completely discount the possibility of the detected PWN 
affecting the pulsar spectrum.

The most remarkable feature about PSR B1800$-$21 was the variation of the low 
frequency spectrum within a timescale of few years. The pulsar spectrum has 
steepened resulting in the turnover frequency shifting from 800 MHz to 1.05 
GHz which changed back to its old shape after a period of few years. The pulsar 
radio emission mechanism is a highly tuned phenomenon \citep{mel00,gil04} and 
it is difficult to envision a scenario where such a spectral change can occur 
without fundamentally altering the emission itself. The more likely possibility
is the change in the intervening medium either in the vicinity of the pulsar or
along the line of sight to the source. The shifting of the spectrum towards 
higher frequencies implies an increase in the relative absorption by the medium 
requiring the pulsar emission to pass through a more dense region. If we take 
into account the pulsar transverse velocity, the structures which will likely 
give rise to the change in the spectrum would be of size $\sim$ 220 AU, for the
changes to happen over roughly three years time span. We have used the distance
to the pulsar to calculate the size of the absorber which will decrease if the 
absorber is located nearer. It was much more difficult to constrain the 
physical parameters of the absorber capable of causing such events without any 
obvious physical structures to guide us. We selected three temperature ranges 
between 100 -- 5000 K to find out the electron densities of the absorber needed
to explain the observed change in turnover frequency. The electron densities 
were much higher between 1000 -- 20000~cm$^{-3}$ as shown in Table \ref{tabcon} 
(Additional absorber). As mentioned earlier these are extreme physical 
conditions in the ISM which justifies the rarity of such detections. One likely
absorber of interest is the ultracompact HII regions which are believed to be 
small photoionized nebulae produced by O and B stars inside the clouds of 
molecular gas and dust in the ISM. These objects are characterized by electron 
densities $>$ 10000~cm$^{-3}$ and temperature around 1000 -- 10000 K and are 
extremely compact in size~$<$ 0.1~pc \citep{woo89}. Another possible candidate 
for such absorbers are the dense filaments seen in SNRs. There have been 
reported compact structures with electron densities $>$ 1000 $^{-3}$ and 
temperatures $>$ 1000 K in the Crab nebulae \citep{san98}. However, it should 
be noted that the Crab nebulae is a special case and it would be difficult to 
find such structures in a standard SNR. In the remainder of this section we 
discuss two events observed in pulsars where changes in pulsar emission over 
short timescales (months to years) have resulted due to possible structures in 
the ISM. 

In light of the discussions carried out above an astrophysical phenomenon that
deserves particular attention is the extreme scattering events (ESEs) in the
ISM. The ESEs are believed to originate due to dense structures in the ISM 
with sizes of the order of AU causing extreme scattering and flux absorption 
when light from background sources pass through them. The ESEs are known to 
span timescales ranging from a few weeks to several years and were first 
detected in flux of compact extragalactic sources like quasars \citep{fie87}. 
These events are pretty rare and particularly harder to detect as they require 
constant surveillance of the flux over periods of several years. In subsequent 
years ESEs have also been seen along the line of sight of some pulsars where 
the events are characterized by the decrease in the pulsar flux and increase in
the dispersion measure lasting months to years before reverting back to the 
original values \citep{les98,mai03,col15}. It is to be noted that the few cases
where multi frequency data are available the flux reduction at lower frequencies
is greater than that at higher frequencies during the ESEs which is a signature
of thermal absorption. The structures giving rise to ESEs are ideally suited to
account for the change in the pulsar spectra over the timescales of years with 
estimated physical sizes in AU scales and densities $\sim$1000~cm$^{-3}$. The 
observed change in the spectrum of PSR B1800$-$21 over a period of few years 
between 2012 and 2014 is a possible extreme scattering event reflected in the 
spectrum of the pulsar.

Another event of interest in connection with change in pulsar flux is the 
phenomenon of echoes associated with the Crab pulsar \citep{bac00,gra00}. An 
echo was seen following the pulsed emission in the pulsar profile and was 
believed to be a result of scattering by ionized clouds far away from the 
pulsar along the line of sight and contained within the Crab nebula. This was 
associated with a reduction in the pulsar flux and increase in the dispersion 
measure lasting several hundred days. Observations were carried out at two 
different frequencies, 327 MHz and 610 MHz, and it is clear that the change in 
the flux was more at the lower frequency \citep{bac00}, which is once again a 
signature of thermal absorption. A number of models have been proposed 
involving refraction and lensing effect by a plasma lens along the line of 
sight \citep{gra11} with varying degree of success. The filamentary structures 
in the outer parts of the Crab nebula with physical dimensions of $\sim$ AU and
electron densities 1000~cm$^{-3}$ are proposed as the most likely cause of 
such events. The rarity of these events also highlights the paucity of such 
structures around a nebula. In a direct extension to our studies, the 
filamentary structures are the ideal candidates to account for the spectral 
changes seen in PSR B1800$-$21 via thermal absorption if they happen to be 
along the line of sight. The presence of a surrounding SNR provides further 
ground for the likelihood of such rare structures around the pulsar. 

To summarize our results, we have carried out detailed observations of the 
pulsar B1800$-$21 at the low radio frequency regime with the objective of 
determining the spectrum. We used two different measurement techniques, the 
phased array and interferometer, and demonstrated the equivalence of these 
methods in estimating the pulsar flux. We confirmed the pulsar spectrum to 
exhibit GPS characteristics \citep[][D15]{kijak2011b}. We reported the first 
instance of a change in the spectrum within a period of years with the turnover
frequency shifting to a higher value before regaining its old form. This is 
most likely due to thermal absorption in compact dense structures along the 
line of sight. We have carried out a detailed statistical fit to the spectrum 
of the pulsar during all the epochs of observation and using basic assumptions 
about three structures in the ISM, filaments in SNR, PWN and HII regions, 
determined the physical parameters responsible for the GPS phenomenon. Based on
the physical arguments we found the HII region to be an unlikely candidate as 
the absorber. The physical conditions in the ISM likely to give rise to the 
detected shift in the turnover frequency was more difficult to constrain. We 
investigated phenomenon like ESEs and echoes in the Crab pulsar where compact 
structures responsible for short term change in pulsar radio emission are 
possible candidates for the spectral change in PSR B1800$-$21. The detection of
the GPS phenomenon in pulsars provide another possibility of probing their 
environments. These objects are usually found around interesting environments 
like PWNe and SNRs and provide additional evidence of association between 
pulsars and these specialized surroundings. The number of pulsars showing GPS 
phenomenon is still very small and we lack information about the physical 
constituents of these environments to distinguish between the absorption at 
different stages of the SNR as well as the PWN. The discovery and 
characterization of the GPS phenomenon in a wider sample of pulsars would 
provided important insights into the physical compositions and the evolution of
their environments.

\section*{Acknowledgments}
This work was supported by grants DEC-2012/05/B/ST9/03924 and 
DEC-2013/09/B/ST9/02177. We thank the anonymous referee for useful comments. We
thank the staff of the GMRT who have made these observations possible. The GMRT
is run by the National Centre for Radio Astrophysics of the Tata Institute of 
Fundamental Research.

\end{document}